\documentclass[aps,showpacs,superscriptaddress,preprint]{revtex4}
\usepackage{amsmath}
\usepackage{amsfonts}
\usepackage{amssymb}
\usepackage{graphicx}
\usepackage[utf8]{inputenc}
\begin{document}
\title{Photon emission by bremsstrahlung and nonlinear Compton scattering in the interaction of ultraintense laser and plasmas}
\author{Feng Wan}
\author{Chong Lv}
\author{Moran Jia}
\author{Haibo Sang}
\affiliation{College of Nuclear Science and Technology, Beijing Normal University, Beijing 100875, China}
\author{Baisong Xie \footnote{Corresponding author. Email address: bsxie@bnu.edu.cn}}
\affiliation{College of Nuclear Science and Technology, Beijing Normal University, Beijing 100875, China}
\affiliation{Beijing Radiation Center, Beijing 100875, China}
\date{\today}
\begin{abstract}
By implementing the bremsstrahlung with Monte Carlo algorithm into the particle-in-cell code, the bremsstrahlung and nonlinear Compton scattering can be studied simultaneously in comparison way in the laser plasma interactions. The simulations are performed for the laser of different intensities interacting with either low-$Z$ or high-$Z$ target. The relative strength of the two photon emission from bremsstrahlung and nonlinear Compton scattering are compared. The result shows that when an ultrastrong intensity laser interacting with a thin and relative high $Z$ target the nonlinear Compton scattering is dominant, however, when the laser intensity $I < 10^{22} \mathrm{W/cm^2}$, the photon emission contributed by bremsstrahlung is comparable to that from nonlinear Compton scattering. In this case the usual ignorable of bremsstrahlung need to be reconsidered.

\end{abstract}
\maketitle

\section{introduction}

The upcoming laser facilities, for example the project of Extreme Light Infrastructure (ELI), promise to provide peak focal intensities over $10^{23} \mathrm{W/cm^2}$ to $10^{25} \mathrm{W/cm^2}$. These laser intensities have the potential ability to unveil the mysteries of quantum vacuum \cite{marklund2006, piazza2012} as well as to reach the limits of attainable intensity of electromagnetic wave \citep{fedotov2010,grismayer2016,zhang2015}. They are conveniently employed to observe the radiation transition from classical to quantum region \citep{zhang2015,jillprl2014} and to provide the efficient sources of $\gamma$-rays and dense anti-matter \citep{burke1997,bamber1999,bell2008,ridgers2013,luo2015,chang2015,zhuxl2015}.

In some recent works \citep{bell2008,kirk2009,brady2012,kirk2013,blackburn2014,ridgers2014}, the laser induced quantum electrodynamics (QED) processes have been investigated extensively. The nonlinear Compton scattering, radiation reaction and pair production by the Breit-Wheeler processes have been implemented into the particle-in-cell (PIC) code like EPOCH \citep{arber2015} and VLPL \citep{jillpop2014} in the laser plasma interaction. These new processes not only provide the new sources for $\gamma$-ray or/and antimatter but also induce some new effects on the classical physical phenomenon such as the electron phase contraction caused by the radiation reaction \citep{lehmann2012,seipt2011}, electron trapping in the near critical density plasma \citep{jillprl2014} and so on.

 On the other hand, for the plasmas with high charge state or/and large atom number $Z$, the processes involving the atom nucleus like the bremsstrahlung ($e + Z \rightarrow e' + \gamma + Z$) or pair production by the Bethe-Heitler processes ($\gamma + Z \rightarrow e^- + e^+ + Z$) may be important in laser plasma interaction \citep{sarri2014,sarri2015,guo2008,liang2015,pike2014}. Usually the radiation through bremsstrahlung is mainly considered in the interaction of electron bunch with high $Z$ target like of copper, tungsten, gold, etc. All these high $Z$ matter is not easily ionized with low intensity lasers, however, with the ultra-intense or/and ultra-relativistic lasers, the high $Z$ plasma can be easily obtained. And at these laser intensities, the bremsstrahlung and nonlinear Compton scattering ($e + n \omega \rightarrow e'+\gamma$) will become the main sources for $\gamma$-ray radiation. Currently most works are engaged in investigating nonlinear Compton scattering or bremsstrahlung individually. A simultaneous comparison study is still lacking, therefore, it is the goal of present research to show the relative importance and photon emission strength for these two mechanisms under different laser intensities.

In this paper we will use PIC code by implementing Monte Carlo (MC) algorithm to study the bremsstrahlung and nonlinear Compton scattering. We will show the relative strength of bremsstrahlung and nonlinear Compton scattering when an ultra-strong laser with intensity ranges from $I=10^{21}$ $\mathrm{W/cm^2}$ to $I=10^{24}$ $\mathrm{W/cm^2}$ irradiating a thin Al or Au target. The $\gamma$-ray distribution and some other characteristics of each mechanisms will be shown in details.

This paper is organized as follows. In Sec. 2, we will review the nonlinear Compton scattering and bremsstrahlung processes and discuss the algorithm of bremsstrahlung by MC implemented to PIC code and the benchmark results. In Sec. 3, we shall discuss the radiation strength of nonlinear Compton scattering and bremsstrahlung with given thin targets and given laser intensities. Summary and discussion has been given in the final section.

\section{Implementation of Bremsstrahlung}

\subsection{comparison of two radiation processes}

The nonlinear Compton scattering is caused by an electron scattering with multiple laser photons $e + n\omega_l \rightarrow e'+\gamma$, which converts several low energy laser photons into a high energy $\gamma$ photon. This mechanism along with the Breit-Wheeler pair production ($ \gamma+ \omega_l \rightarrow e^- + e^+ $) are verified experimentally in the SLAC \citep{bula1996,burke1997,bamber1999}. In the Ref. \citep{kirk2009}, authors implementing the quantum synchrotron radiation \citep{sokolov1968} into the PIC code. The importance of the nonlinear Compton scattering is strongly depending on the laser intensity via the Lorentz invariant $\eta = \gamma \sqrt{(E_\perp+v \times B)^2+(v \cdot E_\parallel )^2}/ E_{cr}$, where $\gamma$ denotes the relativistic Lorentz factor of incoming electrons in laser field, $E_{cr}=m^2c^3/e\hbar$ is the Schwinger critical field, $v$ is the incoming electron velocity. When $\eta$ approaches unity, large numbers of photons will be generated with most probable energy $\epsilon_\gamma \approx 0.44\gamma\eta mc^2$ \citep{bell2008}. Yet a simple cross section is adequate for comparison with other processes.
	
While for the bremsstrahlung, the cross section is highly dependent on the atomic number $Z$ of the target \citep{pike2014,tsai1974}. It is proportional to $\alpha r_e^2 Z^2$, where $\alpha=e^2/\hbar c=1/137$ and $r_e=e^2/mc^2$ are the fine structure constant and the classical electron radius, respectively. Thus the cross section would be increased beside the electron density has been increased when one increases the target $Z$. This will strongly affect the bremsstrahlung emission, for example, for the aluminum target one has $\sigma_b\approx 139\alpha r_e^2$, while for the gold target with $Z=79$ one gets $\sigma_b\approx 6241 \alpha r_e^2$. This fact leads to that at the same laser intensity the increase of atomic number $Z$ will change the relative photon emission strength no matter what is from bremsstrahlung or/and from the nonlinear Compton scattering.

\subsection{simulation method}

Unlike methods used in Refs. \citep{jiang2014,meadowcroft2012,hanus2014}, we will not separate the bremsstrahlung from the laser plasma interaction. To simulate the bremsstrahlung and nonlinear Compton scattering in the laser plasma interaction, MC method has been implemented in our 2D PIC code. The nonlinear Compton scattering part is the same with EPOCH \citep{arber2015} and it has been tested with very good agreement. For the bremsstrahlung part, we have implement the MC Collision model into the code. This part has been tested with the Geant4 code \citep{agostinelli2003}, and the result will be given below.

In the simulation of bremsstrahlung, one of the widely used cross section formula is \citep{tsai1974}
\begin{equation}
\begin{aligned}
\frac{d\sigma_{eZ}}{d\omega}(\omega, y) = & \frac{\alpha r_0^2}{\omega} \lbrace (\frac{4}{3} - \frac{4}{3}y + y^2) \\  & \times [ Z^2(\phi_1 - \frac{4}{3} \mathrm{ln}Z - 4f) + Z(\psi_1 - \frac{8}{3}\mathrm{ln}Z)] \\ & +
\frac{2}{3}(1-y)[Z^2(\phi_1 - \phi_2) + Z(\psi_1 - \psi_2)] \rbrace,
\end{aligned}
\end{equation}
where $y = \hbar \omega / E$ (the energy ratio of the emitted photon to the incident electron), $\phi_{1, 2}$ and $\psi_{1, 2}$ are functions of the screening potential by atomic electrons, and $f$ is the Coulomb correction term. For the high $Z$ target, e.g. $Z > 5$, we shall use Eqs.(3.38-3.41) from Ref. \cite{tsai1974}. For $Z < 5$, the approximated screen functions are not suitable and need to be modified.

Another method which had been used in the code PENELOPE \cite{salvat2009} is the tabulated data from Ref. \cite{seltzer1986}, in which the "scaled" bremsstrahlung differential cross section (DCS) could be transformed to differential cross section by \cite{salvat2009}
\begin{equation}
	\frac{d\sigma_{br}}{d \omega} = \frac{Z^2}{\beta^2} \frac{1}{\omega}\chi(Z, E, y),
\end{equation}
where $\beta = v/c$ is the normalized electron velocity. By integrating the $d \sigma_{br}/d\omega$ with $d\omega$, we can get a tabulated $\sigma_{br}(E, y)$ which could be used for MC simulation.

The DCS for electron and positron is connected as
\begin{equation}
	\frac{d \sigma_{br}^{+}}{d\omega} = F_p(Z, E) \frac{d \sigma_{br}^{-}}{d\omega}
\end{equation}
and the analytical approximation of factor $F_p(Z, E)$ could be found in Ref. \cite{salvat2009}, which shows a good accuracy of about $0.5\%$ in comparison with the Ref. \cite{kim1986}.
	
In our case, the implementation of bremsstrahlung is simply a direct MC collision. For a given incident electron with energy $E$ and velocity $v$, the probability of trigger a bremsstrahlung event is given by
\begin{equation}	
	P_{br} = 1 - e^{n \sigma(E) v \Delta t} = 1 - e^{\Delta s/ \lambda},
\end{equation}
where $n$ denotes the target density, $\Delta t$ is the time interval, $\sigma(E) = \int_{y_{\mathrm{cut}}}^{1} \frac{d \sigma(E, y)}{dy} dy$, $\Delta s = v \Delta t$ and $\lambda = 1 / n\sigma(E)$. Then we will generate a random number $R_1$ to compare with the probability. If $R_1 < P_{br}$, then a bremsstrahlung will be triggered. The photon energy is chosen in the similar way by generate another random number $R_2$, and multiplied with the $\sigma_{br}(E)$ to determine the $\kappa$ through $\sigma(y, E) = \sigma(E) R_2$. Finally, a photon with energy $\hbar \omega = Ey$ and momentum direction $\vec{k}/|k| = \vec{v} / |v|$ will be generated. By choosing a minimum energy of emitted hard photon, we can drop those low energy photons which are not our interests, we can boost our computation. This kind calculation of probability is the same with the method of calculation the random free path \cite{salvat2009}. The implementation of Bethe-Heitler pair production is similar to the bremsstrahlung, and it is not our topic in this article.

\begin{figure}[hbtp]\suppressfloats
\centering
\includegraphics[width=15cm]{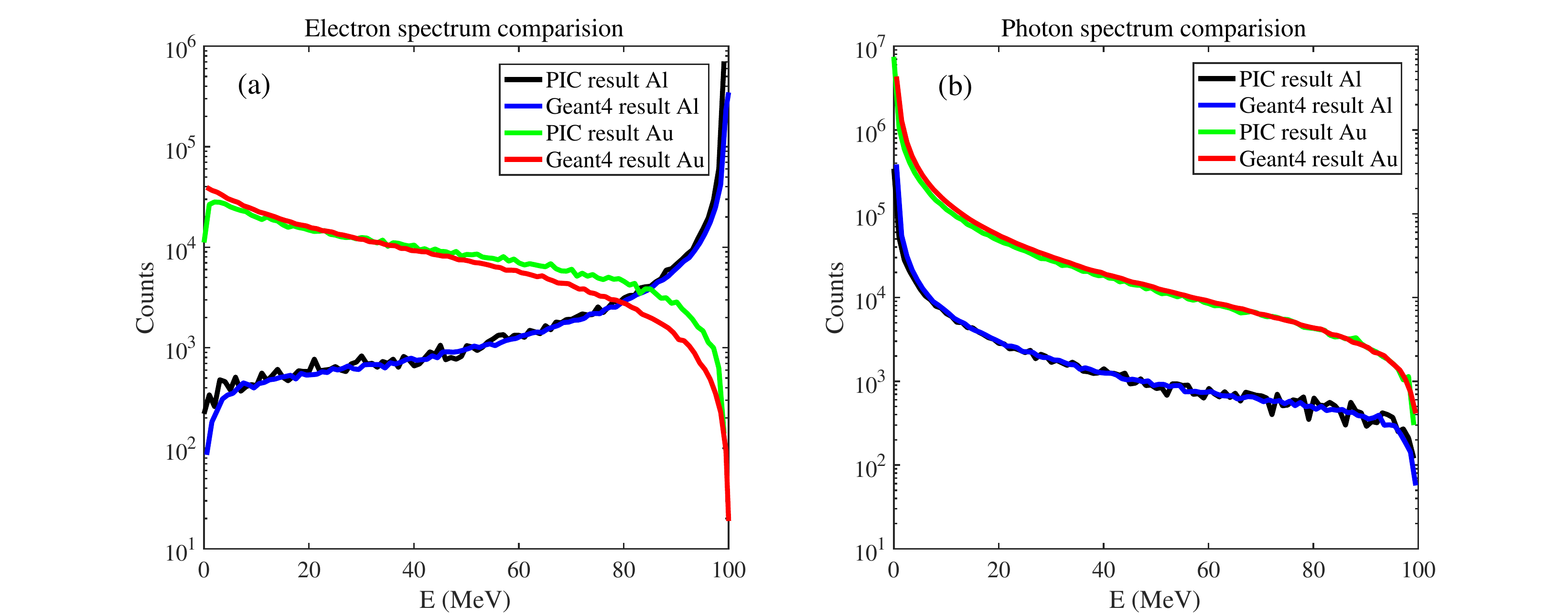}
\caption{(color online). Bremsstrahlung of 100 MeV electrons}
\label{mevbrems}
\end{figure}

\begin{figure}[hbtp]\suppressfloats
\centering
\includegraphics[width=15cm]{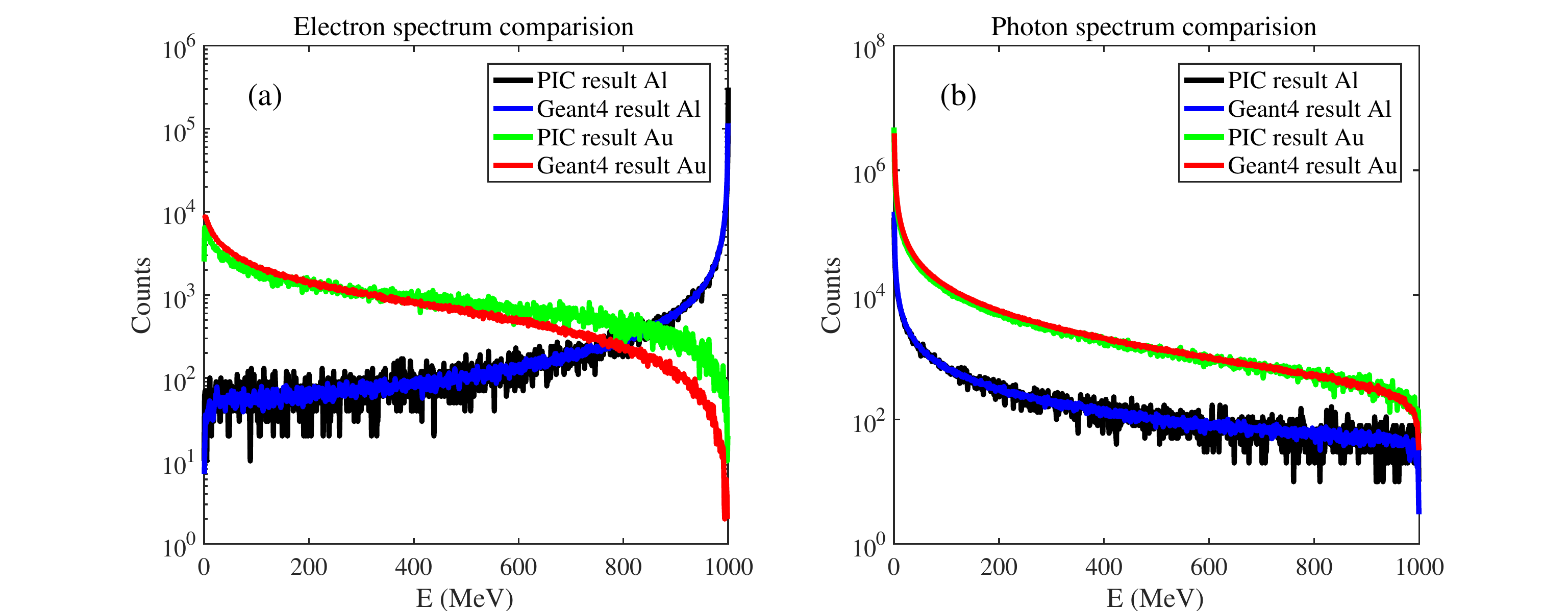}
\caption{(color online). Bremsstrahlung of $1 \mathrm{GeV}$ electrons}
\label{gevbrems}
\end{figure}

Bremsstrahlung emission has been tested with the Geant4 code, which is capable of simulating very comprehensive processes. We have used a 1 GeV and 100 MeV bunch electrons constituted by $10^5$ primaries to collide a 5 mm Au target with $Z=79$, i.e. $\rho=\mathrm{19.3 g/cm^3}$ and a 5 $ \mathrm{mm}$ Al target with $Z=13$, i.e. $\rho=2.7 \mathrm{g/cm^3}$. In the PIC code, we have turn off the field updater and weighting procedure, only particle pusher and bremsstrahlung MC module is enabled. The electron and photon spectra seems to be in good agreement with Geant4 result except a slight higher for electron spectra in the high energy tail. In Fig. \ref{mevbrems} we have plotted the spectra of electron and photon from a  $100\mathrm{MeV}$ electron bunch normally incident onto the aluminum and gold slab and in Fig. \ref{gevbrems} a $1\mathrm{GeV}$ bunch electrons normally incident onto the same target as in Fig. \ref{mevbrems}. And in the following we will use this module to investigate the bremsstrahlung emission in the laser plasma target interaction.

\section{Bremsstrahlung and nonlinear Compton scattering in laser irradiating solid targets}

We have used the aluminum target and gold target to investigate the electron density and atomic number $Z$ effects on the intensity of radiation. Four sets of 2D PIC simulations have been performed to study the relative strength of nonlinear Compton scattering and bremsstrahlung with the laser intensity ranges from $I=10^{21}$ $\mathrm{W/cm^2}$ to $I=10^{24}$ $\mathrm{W/cm^2}$. In all simulations, the lasers are linearly polarized in $y$ direction and propagating along $x$ direction. The temporal profile is set to be constant from 0 to 30 fs, and the spatial profile in $y$ direction is a Gaussian with spot size 1 $\mathrm{\mu m}$. The simulation box covers 6 $\mathrm{\mu m}$ in $x$ and $y$ direction with $1000 \times 1000$ cells for aluminum target and $2000 \times 2000$ cells for gold target, respectively. The plasma target is starting from $x = 2$ $\mathrm{\mu m}$ with 1 $\mathrm{\mu m}$ thickness. The macro particles per cell is $80$ for electrons and $20$ for ions.

All targets has been presumed fully ionized with $n_e \approx 711 n_c$ for Al and $n_e \approx 4177 n_c$ for gold since the ponderomotive $\langle \gamma \rangle \approx 27$ when $I=10^{21}$ $\mathrm{W/cm^2}$, which means that Au could be easily fully ionized in the thin target case \citep{Beiersdorfer2012,mishra2013}. Thus a constant density for two types of targets is assumed in all simulations. Absorbing boundary condition has been used for the laser and particles. Note that in all simulations only photons with energy $\epsilon_\gamma \geq m_ec^2 \approx$ 0.511 MeV is taken into account, while low energy photons are also created but they are dropped to boost the computation.

\begin{figure}[hbtp]\suppressfloats
\centering
\includegraphics[width=15cm]{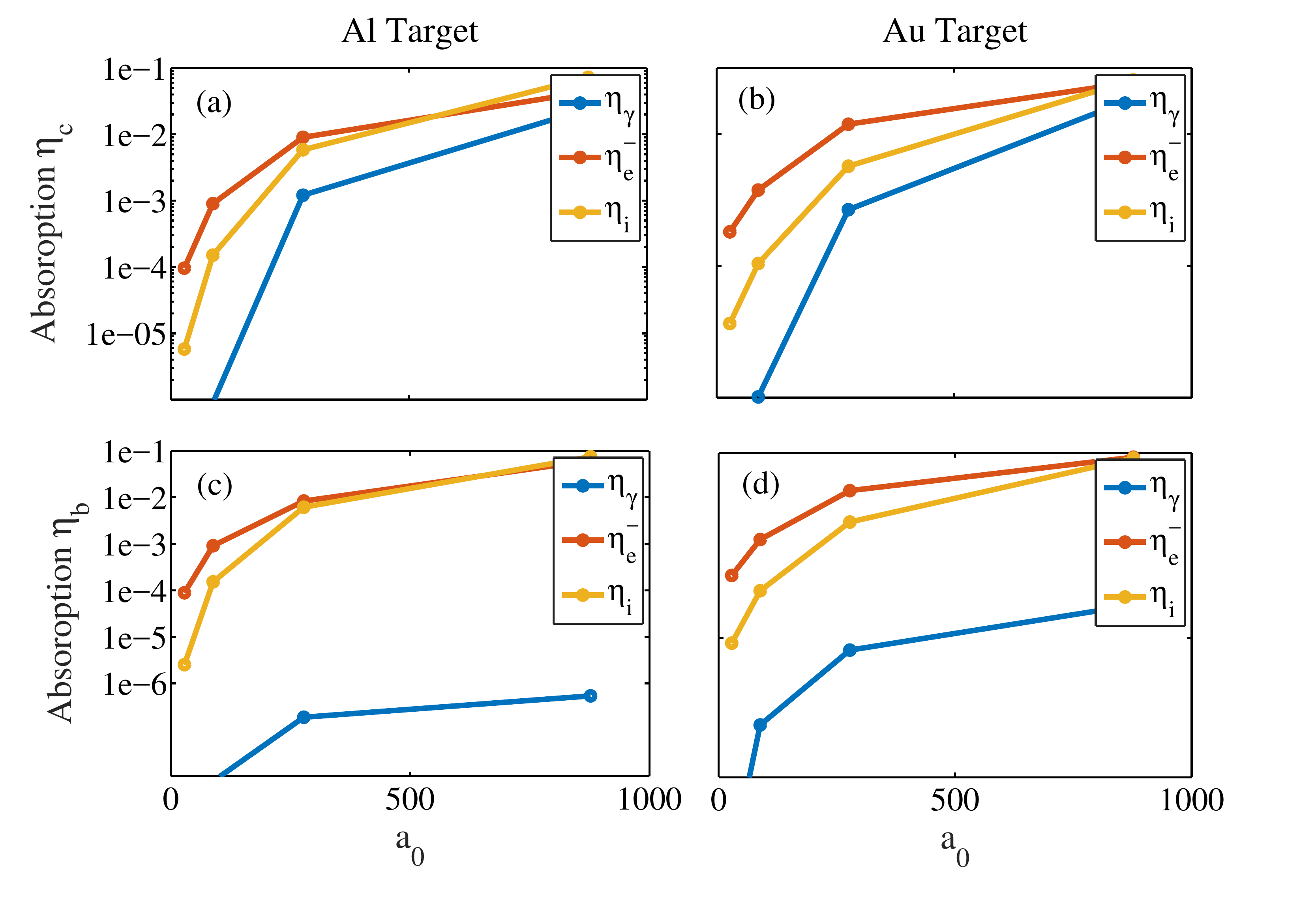}
\caption{(color online). Energy absorption rate, where 'b' denotes the bremsstrahlung, 'c' denotes the nonlinear Compton scattering.}
\label{ekabsorption}
\end{figure}

Now we can give the simulation results for the energy absorption rate of particles and photons. For simplicity we have instead of laser intensity by the normalized vector potential $a_0=0.86\sqrt{I_{18}}$ when the laser wavelength $\lambda=1\mathrm{\mu m}$ is given, where $I_{18}$ means the laser intensity in unit of $10^{18} \mathrm{W/cm^2}$. In Fig. \ref{ekabsorption} (a) and (b), we have plotted the energy partition of both targets with the nonlinear Compton scattering at $t =32$ fs. Since we use a thin target with $l =$ 1 $\mathrm{\mu m}$, so the final result will not be the same with Ref. \citep{jillpop2014}, in which very thick target has been used. In both cases, the absorption rate of $\gamma$-ray increases with the laser intensity. This result agrees with the Ref. \citep{jillpop2014}. But in our case, the interaction time is reduced because not only the piston velocity is larger but also the target is very thin. Thus the absorption rate of $\gamma$-ray will become smaller compared to thick target case \citep{jillpop2014}. Besides, due to lower conversion rate to photons, the electron absorption rate $\eta_e$ is higher than the thick target case and continue to rise with the laser intensity increases. And due to the higher electron density in the Au target case, e.g. $n_{e,Au} \approx 6 n_{e,Al}$, the electrons in the Au target always acquires higher absorption rate than the Al target.

In Fig. \ref{ekabsorption} (c) and (d), we have plotted the energy partition in the bremsstrahlung case. The results of electron and ion trend are similar to that in the nonlinear Compton case, except that electrons absorption is a little higher. Furthermore, bremsstrahlung photons acquires much lower energies than the nonlinear Compton scattering photons, and this difference becomes much more apparent for higher intensities. Thus, this difference is coincident with the difference of electron absorption rate. This indicates that the bremsstrahlung in the cases of very high intensity can be ignored even for high-$Z$ target.

\begin{figure}[hbtp]\suppressfloats
\centering
\includegraphics[width=12cm]{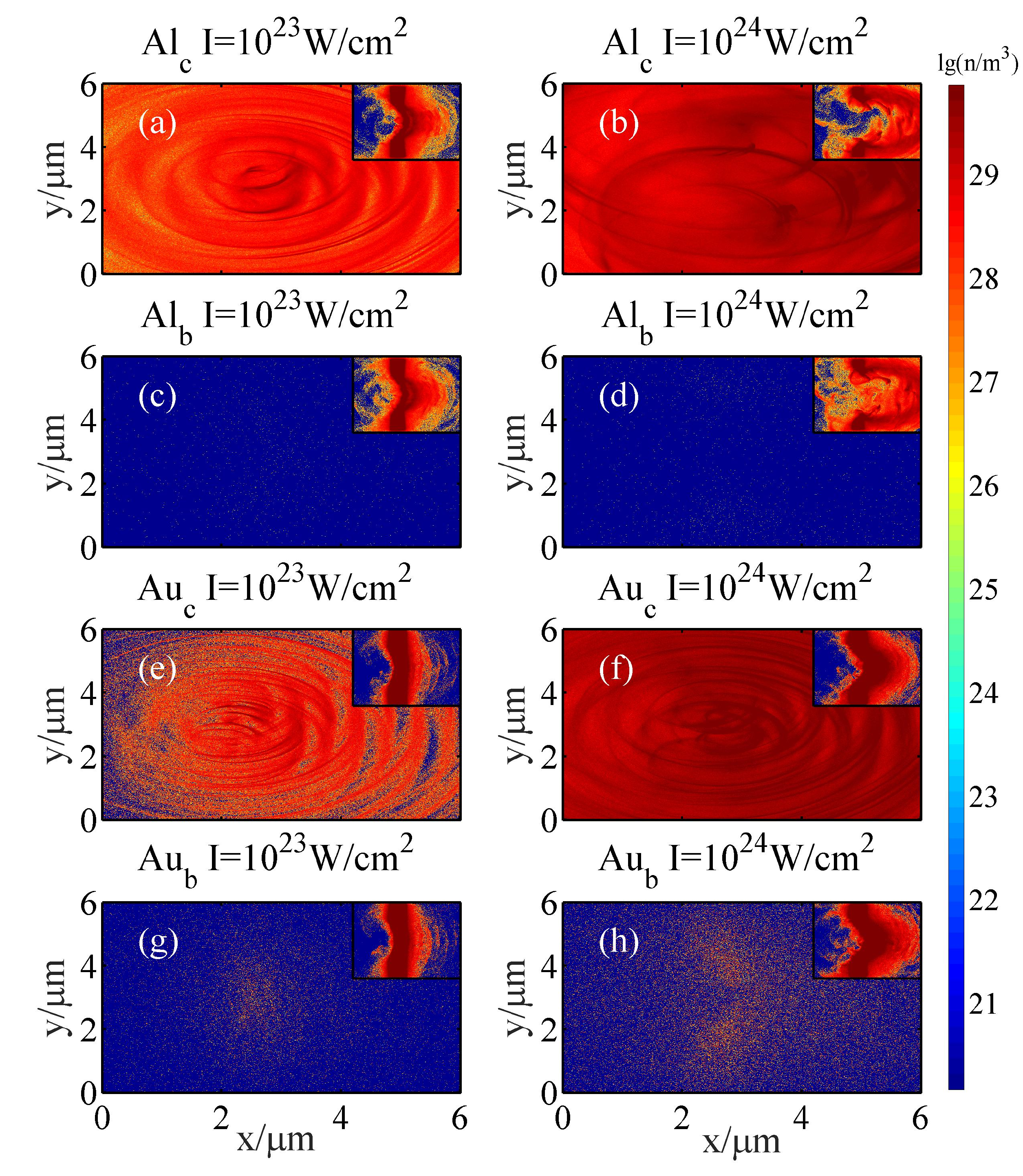}
\caption{(color online). Photon and electron density distribution in log scale, where 'b' denotes the bremsstrahlung, 'c' denotes the nonlinear Compton scattering.}
\label{density}
\end{figure}

In Fig. \ref{density}, the photon distribution of nonlinear Compton scattering and bremsstrahlung has been given at $I=10^{23}$ $\mathrm{W/cm^2}$ and $I=10^{24}$ $\mathrm{W/cm^2}$. The inset is the corresponding electron density. The photon emission in the nonlinear Compton scattering is much stronger than bremsstrahlung for both of Al target and Au target. Besides, the photon density distribution identifies different mechanisms. For the nonlinear Compton scattering, photons are propagating out in a spherical manner due to the same shape of the laser field, see Fig. \ref{density}(a), (b), and (e), (f). While for the bremsstrahlung, photons are focused in the laser plasma interaction zone, see Fig. \ref{density}(c), (d) and (g), (h). Furthermore, due to higher electron density and much larger cross section, the created bremsstrahlung photon density is much higher in the Au target than Al target. There is a little difference between the target deformation for bremsstrahlung and nonlinear Compton scattering. In each target case, number density of electrons residing in the laser front is a little higher for the bremsstrahlung. This may be due to fewer emission events lead to higher electron energies, thus laser was unable to expel these electrons quickly.

\begin{figure}[hbtp]\suppressfloats
\centering
\includegraphics[width=15cm]{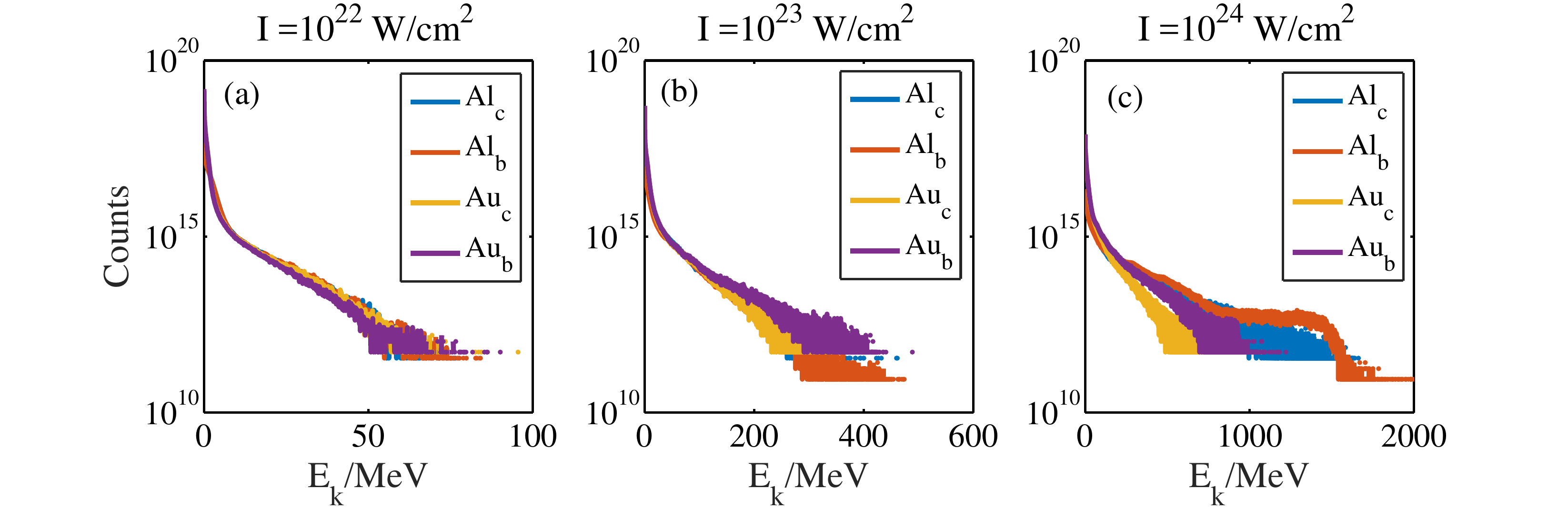}
\caption{(color online). Electron spectra, where 'b' denotes the bremsstrahlung and 'c' denotes the nonlinear Compton scattering.}
\label{electron_spectra}
\end{figure}

In Fig. \ref{electron_spectra} we have plotted the electron energy spectra of two emission mechanisms for different cases. (Note that in the figure here and the next figure the energy of electron as well as photon are both denoted as $E_k$ for convenience.) For the Al target, we can see that in the case of $I=10^{22}$ $\mathrm{W/cm^2}$, electron spectrum is almost the same, see Fig. \ref{electron_spectra} (a). But in the case of $I=10^{23}$ $\mathrm{W/cm^2}$, electrons in the bremsstrahlung case acquires a little higher tail, and this becomes more obvious in the case of $I=10^{24}$ $\mathrm{W/cm^2}$.  Besides, the number of low energy electrons has been reduced compared to lower intensities. By comparing different targets, we can see that electrons in the Al target acquires higher maximum energy than the Au target when $I=10^{24}$ $\mathrm{W/cm^2}$, see Fig. \ref{electron_spectra} (c), but they are almost the same for lower intensities. This is caused by the different target deformation, see Fig. \ref{density}. Since the piston velocity is depending on the target density and laser intensity as $v_{HB}=\Xi/(1+\Xi)$ with $\Xi=I/\rho c^3$ \citep{robinson2009}. Thus the burn out of higher intensities and low $Z$ target is much quicker than lower intensities and high $Z$. If the target has been burn out, electrons in vacuum are oscillating with the laser field without the confinement of plasma space charge field. If the piston has not finished, electron's longitudinal oscillation will be confined by the ion attraction, and the maximum energy will be lower compared with the case in vacuum, which will be given as $\epsilon_{max} \leq \epsilon_l \approx c \Delta p_{laser}=ce/\omega \int_0^\pi E_0 \mathrm{sin}(\phi)d\phi\approx$ 1.8 GeV for $I=10^{24}$ $\mathrm{W/cm^2}$ and $\epsilon_{max} \leq$ 550 MeV for $I=10^{23}$ $\mathrm{W/cm^2}$.

\begin{figure}[hbtp]\suppressfloats
\centering
\includegraphics[width=15cm]{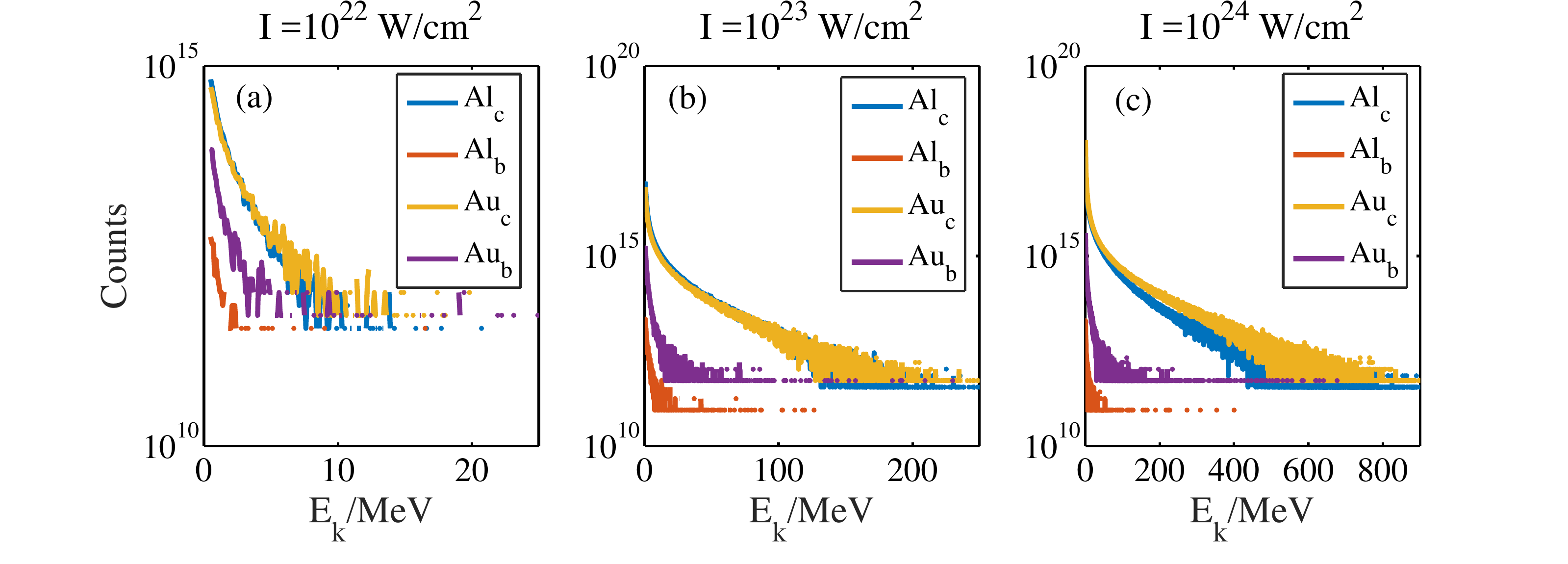}
\caption{(color online). Photon spectra, where 'b' denotes the bremsstrahlung and 'c' denotes the nonlinear Compton scattering.}
\label{photon_spectra}
\end{figure}

In Fig. \ref{photon_spectra}, we have plotted the photon spectra of nonlinear Compton scattering and bremsstrahlung (only photons with $\epsilon_\gamma > 0.511$ MeV are taken into account) for Al and Au targets from different laser intensity. In fact when $I=10^{21}$ $  \mathrm{W/cm^2}$ only the Au target can generate very few photons via the bremsstrahlung which is not shown in the figure. For the nonlinear Compton scattering, the cutoff energy for each case highly depends on the input laser intensity, with 15 MeV for $I=10^{22}$ $\mathrm{W/cm^2}$, 250 MeV for $I=10^{23}$ $\mathrm{W/cm^2}$ and 850 MeV for $I=10^{24}$ $\mathrm{W/cm^2}$, while they are independent of the target type. Besides, even though the electron density of Au target is much larger than the Al target, the photon spectra is almost the same for each kind of intensity, which is the direct result of nearly the same electron spectra.

The photon spectra by bremsstrahlung is quite different from that by nonlinear Compton scattering. Not only the created number is much less than the later mechanisms, also the cut-off energy is much smaller. Since the bremsstrahlung is not directly depending on the laser intensity, the cut-off will be depending on the electron energy and target density etc. As expected, the number of bremsstrahlung photons from the Au target is much larger than that from the Al target, however, it is still much smaller than those by the nonlinear Compton scattering from the Au target.

\begin{figure}[hbtp]\suppressfloats
\centering
\includegraphics[width=15cm]{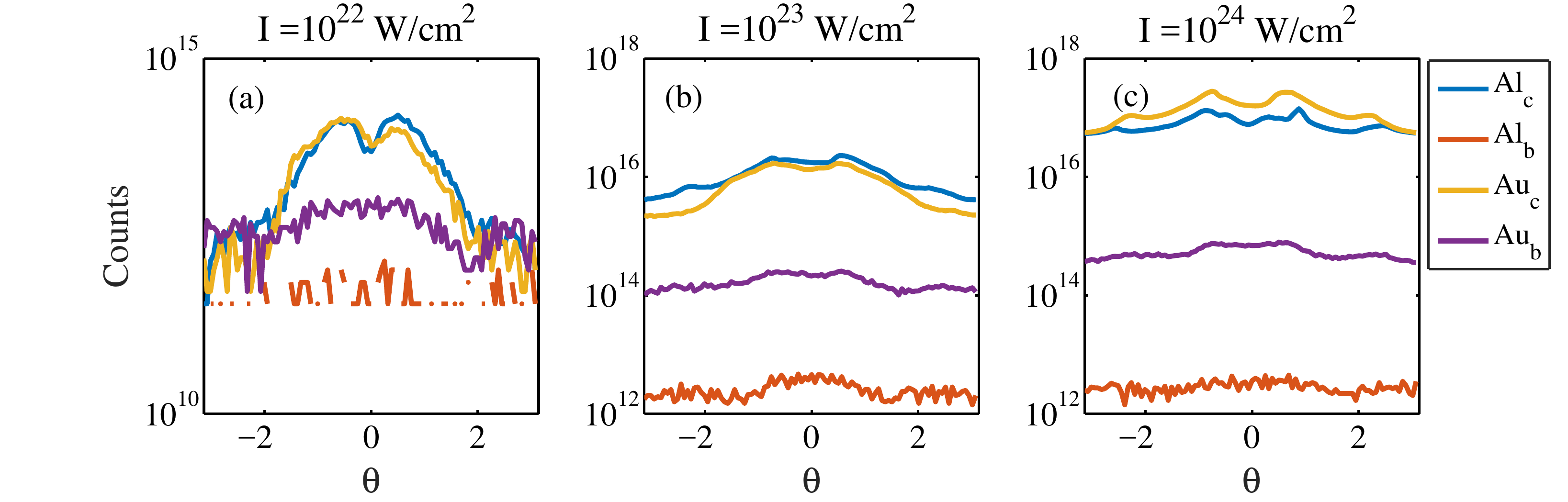}
\caption{(color online). Photon angular distribution, where 'b' denotes the bremsstrahlung and 'c' denotes the nonlinear Compton scattering.}
\label{photon_angle}
\end{figure}

In Fig. \ref{photon_angle}, the angular distribution of photons is exhibited by two kinds of radiation from two targets. We can see that the angular distribution from two kinds of radiation mechanisms is quite different. In the nonlinear Compton scattering case, most photons are focused to the laser polarization nearby with a little deviation. With the increase of laser intensity, the radiation intensity increases, but there are still two peak angles along the polarization direction. In the bremsstrahlung case, due to much smaller yields, the angular distribution is very rough. But we can still see that there is a plateau from $-1<\theta<1$, which is quite different from the nonlinear Compton scattering. The difference can be understood from two different mechanisms. As we demonstrated above, the strength of nonlinear Compton scattering highly depends on the quantum parameter $\eta\approx \gamma E/E_{cr}$. First the electrons in the high field region will be easily accelerated to high $\eta$ which leads to a larger probability of nonlinear Compton scattering. Second the $\eta$ also depends on the polarization. In the bremsstrahlung case, however, the radiation depends on the electron energy and target density which is not directly affected by the polarization so that the angular distribution of bremsstrahlung is not very sensitive to the polarization. For the Au target case, the electron angular is very similar to the Al case. The photon angular distribution is much smooth compared to the Al target case due to larger cross section.

\section{summary and discussion}

In summary, we have implemented a MC collision method into PIC to simulate the bremsstrahlung in the laser plasma interactions. By simulating the laser irradiating Al target and Au target with different laser intensity, we have obtained the relative strength of radiation for each mechanism. From the comparison, we can see that when the laser intensity $I \leq 10^{22} \mathrm{W/cm^2}$, bremsstrahlung is still very strong compared with the nonlinear Compton scattering in the laser high-$Z$ target interaction. And when $I \leq 10^{21} \mathrm{W/cm^2}$, this photon channel dominates the photon emission than the nonlinear Compton scattering for high $Z$ target like Au. Thus this kind of energy conversion may need to be taken into account in seeking the accurate simulation and analytical solutions.

Our research confirms that the bremsstrahlung in the interaction of ultra-intense laser with low $Z$ plasma can be ignored. And for laser intensity $I \geq 10^{22} \mathrm{W/cm^2}$, even for high $Z$ target, the nonlinear Compton scattering is still the dominant radiation channel. Besides, the photon density distribution could be a signature to distinguish the main radiation channel. Since using the MC collision method is very time consuming for large number system, the so called Null-Collision method may be one of the potential methods to reduce computation resources. Besides, to confirm the experimental thick target case, a $\mathrm{mm}$ scale target should be considered, which is beyond the scope of present study and is worthy to be researched in the future work.

\section{acknowledgements}

Authors are grateful to Prof. H Wang for helpful discussions on the implementation of MC algorithm. This work was supported by the National Natural Science Foundation of China (NSFC) under Grants No. 11475026 and No. 11305010. The computation was carried out at the High Performance Scientific Computing Center (HSCC) of the Beijing Normal University. The authors are particularly grateful to CFSA at University of Warwick for allowing us to use the EPOCH.

\bibliographystyle{unsrt}

\end{document}